\documentclass{PoS}

\title{Multipoint reweighting method and beta functions for the calculation of QCD equation of state}

\ShortTitle{Multipoint reweighting method and beta functions for the calculation of QCD equation of state}

\author{\speaker{Ryo Iwami}$^1$, S. Ejiri$^2$, K.~Kanaya$^3$, 
Y.~Nakagawa$^1$, T.~Umeda$^4$, D.~Yamamoto$^1$ (WHOT-QCD Collaboration)  \\ \\
$^1$Graduate School of Science and Technology, Niigata University, Niigata 950-2181, Japan \\
      \ \ \ E-mail: \email{iwami@muse.sc.niigata-u.ac.jp} \\
$^2$Department of Physics, Niigata University, Niigata 950-2181, Japan \\
$^3$Graduate School of Pure and Applied Sciences, University of Tsukuba, 
Tsukuba, Ibaraki 305-8571, Japan \\
$^4$Graduate School of Education, Hiroshima University, Hiroshima 739-8524, Japan 
}

\abstract{
We study a reweighting method aiming at numerical studies of QCD at
finite density, in which the conventional Monte-Carlo method cannot be applied
directly. One of the most important problems in the reweighting
method is the overlap problem.
To solve it, we propose to perform
simulations at several simulation points and combine their results 
in the data analyses. 
In this report, we introduce this multipoint reweighting method and test
if the method works well by measuring histograms of physical quantities.
Using this method, we calculate the meson masses as continuous
functions of the gauge coupling $\beta$ and the hopping parameters
$\kappa$ in QCD at zero density.
We then determine lines of constant physics in
the $(\beta, \kappa)$ space and evaluate the derivatives of the
lattice spacing with respect to $\beta$ and $\kappa$ along
the lines of constant physics (inverse of the beta functions), which are needed in a calculation
of the equation of state.
}

\FullConference{The 32nd International Symposium on Lattice Field Theory\\
                 23-28 June, 2014\\
                 Columbia University New York, NY}

\begin{document}

%%%%%%%%%%%%%%%%%%%%%%%%%%%%%%%%%%%%%%%%%%
\section{Introduction}
\label{sec:introduction}

In a study of QCD phase diagram at finite temperature $(T)$ and density, the complex quark determinant causes a serious problem in numerical simulations.
The reweighting method is commonly used to avoid this problem in the low density region. 
However, when we increase the chemical potential $(\mu)$, the sign problem and the overlap problem becomes severe.
In this report, we focus on the overlap problem. 
The overlap problem is expected to be milder if one changes a couple of parameters at the same time. 
For example, WHOT-QCD collaboration investigated the phase structure of $N_{\rm f}$-flavor QCD in the heavy-quark region, 
and found that the system at large quark masses is controlled by only two combinations of parameters, $\beta +48 \sum_{f=1}^{N_{\rm f}} \kappa_f^4$ and 
$\sum_{f=1}^{N_{\rm f}} \kappa_f^{N_t} \cosh (\mu_f/T)$ on an $N_s^3 \times N_t$ lattice, 
where $\beta=6/g^2$, and $\kappa_f$ and $\mu_f$ are the hopping parameter and chemical potential of $f^{\rm th}$-flavor, respectively \cite{whot14}.
This means that, when one changes the parameters with keeping these combinations constant, the system does not change, thus the overlap problem does not arise.
We expect such combinations of parameters exist also in the light-quark region.

In that study in the heavy-quark region \cite{whot14}, the multipoint reweighting method \cite{FS89} for $\beta$ played an important role: 
Combining data obtained at different $\beta$, we could calculate the effective potential in a wide range of the observable values, that was mandatory in a reliable evaluation of the transition point.
In the present paper, we extend the multipoint reweighting method in the multi-parameter space of $\beta$ and $\kappa_f$ to overcome the overlap problem in the light-quark region.

In the next section, the multipoint reweighting method is introduced, and, in Sec.\ref{sec:overlap}, we examine the overlap problem performing numerical simulations in two-flavor QCD. 
We then calculate the meson masses, the lines of constant physics, and the derivatives of the lattice spacing with respect to $\beta$ and $\kappa$ along the lines of constant physics in Sec.~\ref{sec:beta-func}. 
The section \ref{sec:conclusion} is for the conclusions.

%%%%%%%%%%%%%%%%%%%%%%%%%%%%%%%%%%%%%%%%%%
\section{Multipoint reweighting method}
\label{sec:method}

We extend the multipoint reweighting method to the multi-parameter space such that both gauge and quark parameters are changed simultaneously.
Let us first define a histogram for a set of physical quantities 
$X = (X_1,X_2,\cdots)$ as 
\begin{eqnarray}
w(X; \beta, \vec\kappa, \vec\mu) 
%&=& \int {\cal D} U {\cal D} \psi {\cal D} \bar{\psi} \ \prod_i \delta(X_i - \hat{X}_i) \ e^{- S_q - S_g} \nonumber \\
&=& \int {\cal D} U \ \prod_i \delta(X_i - \hat{X}_i) \ 
e^{-S_g}\ \prod_{f=1}^{N_{\rm f}} \det M(\kappa_f, \mu_f) .
\label{eq:dist}
\end{eqnarray}
where $S_g$ is the gauge action, $M$ is the quark matrix and 
$\hat{X} = (\hat{X}_1,\hat{X}_2,\cdots)$ is the operators for $X$.
We denote $S \equiv S_g - \sum_f \ln \det M$.
The coupling parameters of the theory are $\beta, 
\vec{\kappa}=(\kappa_1, \kappa_2, \cdots, \kappa_{N_{\rm f}}),$ and 
$\vec{\mu}=(\mu_1, \mu_2, \cdots, \mu_{N_{\rm f}})$. 
Then, the partition function is given by
%\begin{equation}
$ Z(\beta,\vec\kappa,\vec\mu)  = \int\! w(X;\beta,\vec\kappa,\vec\mu) \, dX $
%\end{equation}
with $dX = \prod_i dX_i$, 
and the probability distribution function of $X$ is given by $Z^{-1} w(X;\beta,\vec\kappa,\vec\mu)$. 
The expectation value of an operator ${\cal O} [\hat{X}]$ which is written in terms of $\hat{X}$ is calculated by 
\begin{eqnarray}
\langle {\cal O }[\hat{X}] \rangle_{(\beta, \vec\kappa,\vec\mu)} = \frac{1}{Z(\beta,\vec\kappa,\vec\mu) } 
\int\! {\cal O} [\vec{X}] \, w(\vec{X}; \beta, \vec\kappa, \vec\mu) \, dX .
\label{eq:expop}
\end{eqnarray}
%For convenience, we define the effective potential as
%\begin{equation}
%V_{\rm eff}(\vec{X}; \beta, \vec\kappa, \vec\mu) = -\ln w(X; \beta, \vec\kappa, \vec\mu) .
%\end{equation}
%To avoid complicated notations, 
For simplicity, we denote the set of coupling parameters $(\beta, \vec\kappa, \vec\mu)$ as $\beta$ in the following of this section.

We perform a simulation at $\beta_0$ and calculate the histogram at $\beta$.
%Then, let us discuss the coupling parameter-dependence in the histogram.
We fix three quantities $X,$ $S(\beta) \equiv S$ 
and $S(\beta_0) \equiv S_0$ to construct the histogram, 
where $S$ is the value of the action with the coupling parameters $\beta$ on the configuration generated at $\beta_0$.
From Eq.(\ref{eq:dist}), we find
\begin{equation}
w(X, S,S_0; \beta) 
= e^{-(S-S_0)} \, w(X, S, S_0; \beta_0).
\label{eq:chbeta}
\end{equation}
%For example, in the case of the quenched QCD with 
%$S=-\beta (c_0 \sum W^{(1 \times 1)}_{\mu \nu} + c_1 \sum W^{(1 \times 1)}_{\mu \nu}),$ 
%the histogram of$ P \equiv c_0 \sum W^{(1 \times 1)}_{\mu \nu} + c_1 \sum W^{(1 \times 1)}$ changes as 
%\begin{equation}
%w(P; \beta_1, \vec\kappa_1) 
%= e^{(\beta_1 - \beta_0)P} w(P; \beta_0, \vec\kappa_0) ,
%\end{equation}
%where $ w(P; \beta, \vec\kappa)= w(P, S_1, S_0; \beta, \vec\kappa)$
%because $S= \beta P$, i.e. the action is fixed if $P$ is fixed.
%The effective potential changes as
%$V_{\rm eff}(P; \beta_1, \vec\kappa_1) 
%= V_{\rm eff}(P; \beta_0, \vec\kappa_0) -(\beta_1 - \beta_0)P$.
%When the distribution of $P$ is a Gaussian function centered at $P_0$ like 
%$ -ln W(P) = V_{\rm eff} = \alpha (P - P_0)$ with an appropriate constant $\alpha$,
%The peak of the distribution, i.e. $dV_{\rm eff}/dP=0$, is shifted from 
%$P_0$ to $P_0 + (\beta_1 - \beta_0)/(2 \alpha)$ under the change from $\beta_0$ 
%to $\beta_1$.
%The information of the histogram around the peak is important for the calculation of Eq.~(\ref{ eq:expop }). 
%However, when $\beta_1 - \beta_0$ is large, the peak position goes out of 
%the measureable region of the histogram, that is the overlap problem.
The histogram of $X$ is given by 
%\begin{eqnarray}
$
w(X; \beta) = 
\int w(X, S, S_0; \beta) \, dS \,dS_0 
$.
%\label{eq:wint}
%\end{eqnarray}
%from $w(X, S, S_0; \beta, \vec\kappa)$.

When $X$ has large correlation with the difference of the actions, $S - S_0$, 
the peak position of the distribution may change appreciably, 
causing the overlap problem.
(See Sec.~\ref{sec:overlap}.)
To overcome the overlap problem and to obtain $w$ and $V_{\rm eff}$ that are reliable 
in a wide range of $X$, we extend the reweighing formulas to combine data obtained 
at different simulation points \cite{whot14,FS89} for the case including the fermion action. 
%We combine data at different values of $\beta$ and $\vec\kappa$, and suppress 
%the arguments $\vec\kappa$ for simplicity of the notations. 

We combine a set of $N_{\rm sp}$ simulations performed at $\beta_i$ with the number of 
configurations $N_i$ where $i=1, \cdots , N_{\rm sp}$.
Using Eq.~(\ref{eq:chbeta}), the probability distribution function at $\beta_i$ is related to that at $\beta$ as  
\begin{eqnarray}
Z^{-1}(\beta_i) \,w(X,S, \vec{S}; \beta_i) = Z^{-1}(\beta_i) \, e^{-(S_i -S)} \,w(X,S, \vec{S}; \beta) .
\end{eqnarray}
%where $S_i=S(\beta_i, \vec{\kappa})$.
Summing up these probability distribution functions with the weight $N_i$, 
\begin{eqnarray}
\sum_{i=1}^{N_{\rm sp}} N_i \, Z^{-1}(\beta_i) \, w(X, S, \vec{S}; \beta_i) 
= e^{S} 
\sum_{i=1}^{N_{\rm sp}} N_i \, Z^{-1}(\beta_i) \, e^{-S_i}  \, w(X, S, \vec{S}; \beta), 
\label{eq:sum1}
\end{eqnarray}
we obtain 
\begin{eqnarray}
w(X,S, \vec{S}; \beta)= G(S, \vec{S};\beta,\vec\beta) \,
\sum_{i=1}^{N_{\rm sp}} N_i \, Z^{-1}(\beta_i) \, w(X,S, \vec{S}; \beta_i) 
\end{eqnarray}
where $\vec\beta=(\beta_1,\cdots,\beta_{N_{\rm sp}})$ and 
\begin{eqnarray}
G(S, \vec{S};\beta,\vec\beta)=\frac{ e^{-S}}{
\sum_{i=1}^{N_{\rm sp}} N_i \, e^{-S_i} Z^{-1}(\beta_i)} .
\end{eqnarray}
Note that the left-hand side of Eq.~(\ref{eq:sum1}) gives a naive histogram using all the configurations disregarding the difference in the simulation parameter $(\beta, \vec{\kappa}).$
The histogram $w(X,S, \vec{S};\beta)$ at $(\beta, \vec{\kappa})$ is given by multiplying 
$G(S, \vec{S};\beta,\vec\beta)$ to the naive histogram.

The partition function is given by
\begin{eqnarray}
Z(\beta)= \sum_{i=1}^{N_{\rm sp}} N_i \int G(S,\vec{S};\beta,\vec\beta) \, Z^{-1}(\beta_i) \, w(X,S, \vec{S}; \beta_i) \,dXdSd\vec{S}
=\sum_{i=1}^{N_{\rm sp}} N_i \left\langle G(S,\vec{S};\beta,\vec\beta) \right\rangle_{\!(\beta_i)}.
\end{eqnarray}
The right-hand side is just the naive sum of $G(S,\vec{S};\beta,\vec\beta)$ obtained on all the configurations.
The partition function at $\beta_i$ can be determined, up to an overall factor, by the consistency relations, 
\begin{eqnarray}
Z(\beta_i) 
=\sum_{k=1}^{N_{\rm sp}} N_k \left\langle G(S,\vec{S};\beta_i,\vec\beta) \right\rangle_{\! (\beta_k)}
=\sum_{k=1}^{N_{\rm sp}} N_k \left\langle 
\frac{e^{-S_i}}{
\sum_{j=1}^{N_{\rm sp}} N_j e^{-S_j} Z^{-1}(\beta_j)} \right\rangle_{\! (\beta_k)}
\end{eqnarray}
for $i=1,\cdots,N_{\rm sp}$. 
Denoting $f_i=-\ln Z(\beta_i)$, these equations can be rewritten by
%\begin{eqnarray}
\\ $
1 = \sum_{k=1}^{N_{\rm sp}} N_k \left\langle
\left( \sum_{j=1}^{N_{\rm sp}} N_j \exp[S_i -S_j - f_i +f_j] \right)^{-1} 
\right\rangle_{(\beta_k)}.
$ 
%\hspace{5mm}
%i=1,\cdots,N_{\rm sp}.
%\label{eq:consis}
%\end{eqnarray}
Starting from appropriate initial values of $f_i$, we solve these equations numerically by an iterative method. 
Note that, in this calculation, one of $f_i$ must be fixed to remove the ambiguity corresponding to the undetermined overall factor.

The expectation value of an operator $\hat{X}$ at $\beta$ can be evaluated as
\begin{eqnarray}
\langle \hat{X} \rangle_{(\beta)} 
\;=\; \frac{1}{Z(\beta)} \int X \, w(X,S, \vec{S}; \beta) \, dS \,d\vec{S}
\;=\; \frac{1}{Z(\beta)} \sum_{i=1}^{N_{\rm sp}} N_i \left\langle \hat{X} \, G(S,\vec{S};\beta,\vec\beta) \right\rangle_{\!(\beta_i)},
\label{eq:multibeta}
\end{eqnarray}
and the histogram of $X$ is obtained by
\begin{eqnarray}
w(X;\beta) 
\;=\; \sum_{i=1}^{N_{\rm sp}} N_i \left\langle \delta(X- \hat{X}) 
\, G(S,\vec{S};\beta,\vec\beta) \right\rangle_{\!(\beta_i)},
\label{eq:multihis}
\end{eqnarray}
Note that, again, $\sum_{i=1}^{N_{\rm sp}} N_i \left\langle \hat{X} G \right\rangle_{(\beta_i)}$ 
in the right-hand side is just the naive sum of $X G$ over all the configurations disregarding the difference in the simulation point $\beta_i$.

%%%%%%%%%%%%%%%%%%%%%%%%%%%%%%%%%%%%%%%%%%
\section{Overlap problem and histograms}
\label{sec:overlap}

To study if the multipoint reweighting method is useful in avoiding the overlap problem, we perform simulations of QCD with degenerate 2 flavors of clover-improved Wilson quarks coupled with RG-improved Iwasaki glue, at $\mu=0$.
%The Iwasaki gauge and clover Wilson fermion actions are used.
The improvement parameters of the action are the same as those adopted in Ref.~\cite{whot10}.
The simulations are carried out on an $8^4$ lattice at 9 simulation points 
(3 $\beta$'s $\times$ 3 $\kappa$'s) for the test study in this section, and 
on a $16^4$ lattice at 30 points (6 $\beta$'s $\times$ 5 $\kappa$'s) in Sec.~\ref{sec:beta-func}.
The number of configurations for the measurement is 200 at each simulation point. 

%\paragraph{Calculation of the fermion determinant}
The action $S$ required in the reweighting method is composed by Wilson loops and 
$\ln \det M$.
%the logarithm of the fermion determinant.
The calculation of the fermion part requires large computational cost.
In this study, we evaluate 
%the fremion determinant 
$\det M$ by measuring the first and 
second derivatives of $\ln \det M$ at several $\kappa_i$'s, where $i=1, 2, \cdots$, on each configuration,  
and interpolate the determinant between $\kappa_i$ and $\kappa_{i+1}$ 
assuming a quadratic function, 
$\ln \det M(\kappa) = \ln \det M(\kappa_i) +C_1 (\kappa -\kappa_i) +C_3 (\kappa -\kappa_i)^2 
+C_3 (\kappa -\kappa_i)^3 +C_4 (\kappa -\kappa_i)^4$.
The coefficients $C_a$ are determined such that the first and second derivatives 
are consistent with the measured values at $\kappa_i$ and $\kappa_{i+1}$.
The clover term with the coefficient $c_{SW}$ is also evaluated in terms of its derivatives. 
Here, because $c_{SW}$ depends on $\beta$ in our choice,   
it affects the $\beta$-dependence of observables.
However, we find that the effects from $c_{SW}$ are very small in the range of $\beta$ relevant in our study.
We thus approximate the $\beta$-dependence of the action by a linear function, i.e. 
$\ln \det M(\beta, \vec{\kappa})= \ln \det M(\beta_0, \vec{\kappa}) 
+ (\beta - \beta_0) (d c_{SW}/d \beta) (\partial \ln \det M / \partial c_{SW}) (\beta_0, \vec{\kappa})$ \cite{CswBeta}.
Note that the overall constant of $\ln \det M$ is not needed.

\begin{figure}[t]
\begin{center}
\centerline{
\includegraphics[width=54mm]{improve_8888_MKREW_PoS_P.eps}
\hspace{-7mm}
\includegraphics[width=54mm]{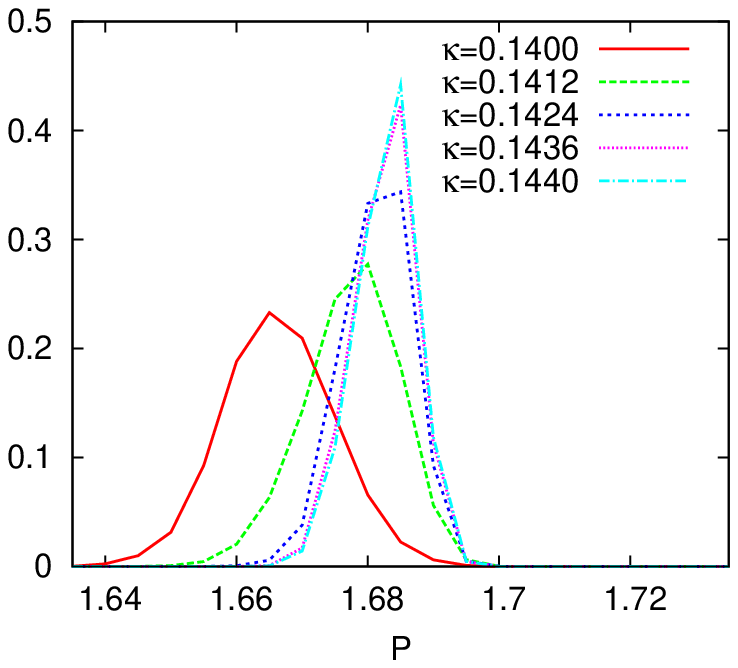}
\hspace{-8mm}
\includegraphics[width=54mm]{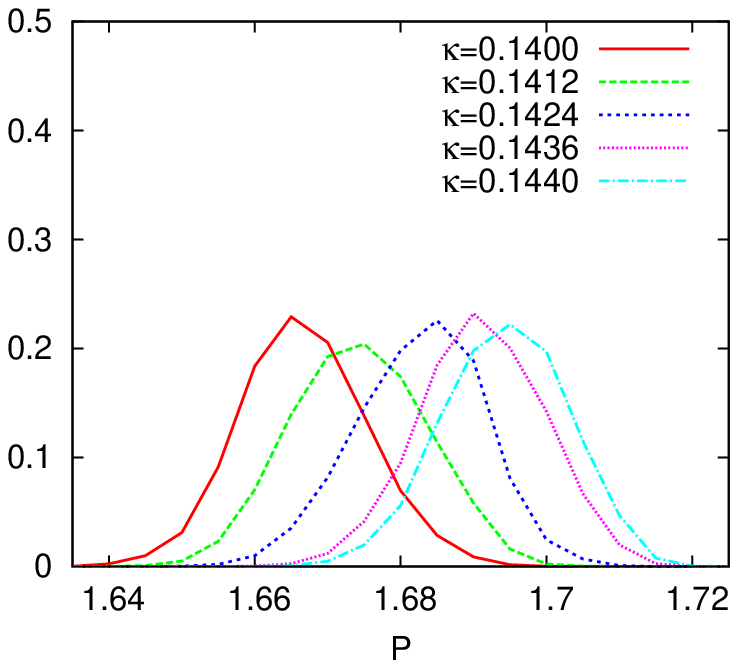}
}
%\vspace{0mm}
\caption{Left: The expectation value of $P \equiv c_0 W^{1 \times 1} + 2 c_1 W^{1 \times 2}$ as a function of $\kappa$ at $\beta=1.825$.
Middle: The histogram of $P$ at various $\kappa$'s obtained by the naive reweighting method using the configurations at $\kappa=0.140$.
Right: The histogram of $P$ by the multipoint reweighting method.}
\label{fig1}
\end{center}
\end{figure}

\begin{figure}[t]
\begin{center}
\vspace{-13mm}
\centerline{
\includegraphics[width=87mm]{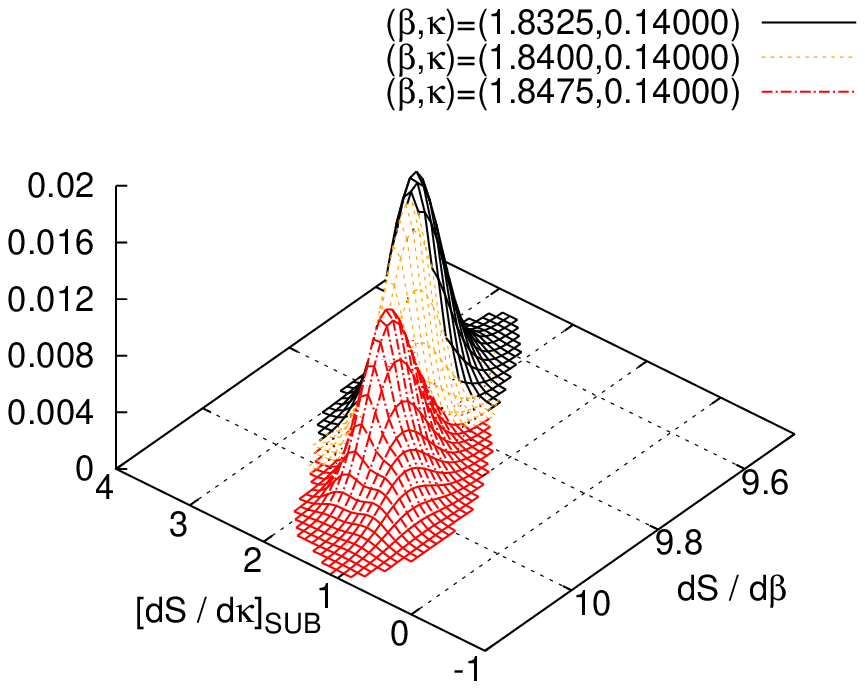}
\hspace{-13mm}
\includegraphics[width=87mm]{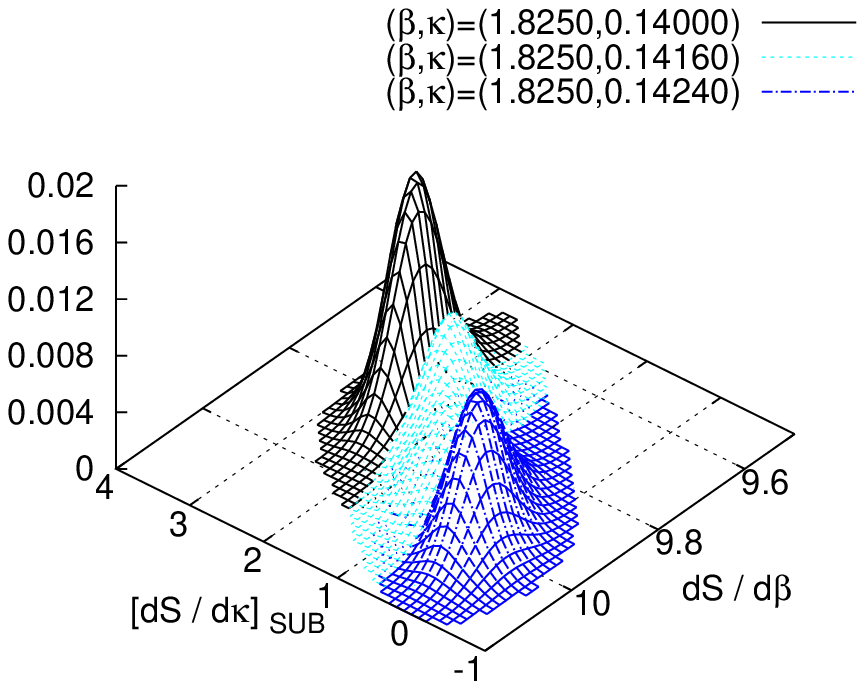}
}
\vspace{-5mm}
\caption{The $\beta$-dependence (left) and $\kappa$-dependence (right) of the histogram 
for $N_{\rm site}^{-1} (\partial S/ \partial \beta)$ and 
$N_{\rm site}^{-1} [\partial S/ \partial \kappa]_{\rm SUB} \equiv 
N_{\rm site}^{-1} [\partial S/ \partial \kappa 
-(288 N_{\rm f} \kappa^4/c_0) (\partial S/ \partial \beta)]$, where $N_{\rm site}=8^4.$}
\label{fig2}
\end{center}
\vspace{-8mm}
\end{figure}

In the left panel of Fig.~\ref{fig1}, we show the results for the improved plaquette
$P=c_0 W^{1 \times 1} + 2c_1 W^{1 \times 2}$ of the Iwasaki action at $\beta=1.825$, 
where $c_1=-0.331$, $c_0=1-8c_1$, and $W^{i \times j}$ is the $(i \times j)$ Wilson loop.
Black dots represent the expectation values of $P$ at the three simulation points without reweighting.
Blue, green and purple curves are the results of the naive reweighting method 
using the data at $\kappa=0.1400$, 0.1425, and 0.1440, respectively. 
Each result of the naive reweighting method is reliable around the corresponding simulation point, 
but fails reproducing far away simulation results.
The reason can be easily understood by consulting the histogram of $P$:
Red curve in the middle panel of Fig.~\ref{fig1} is the original histogram at 
$(\beta, \kappa)=(1.825, 0.140)$, and green, blue, magenta and light blue curves are the histograms at $\kappa=0.1412$, 0.1424, 0.1436, and 0.144, respectively, estimated by the naive reweighting method Eq.~(\ref{eq:chbeta}) using the data at $\kappa=0.140$.
Because the histograms at $\kappa$ other than the simulation point is calculated as the product of 
the reweighting factor and the original histogram, 
the histograms are not reliable out of the range of the original distribution.
In fact, the value of $P$ distributes between $1.64$ and $1.69$. Even when $\kappa$ is 
changed by the reweighting method, the upper and lower bounds of the distribution does not change in Fig.~\ref{fig1} (middle).
Since the expectation value is approximately the peak position of the histogram, 
the expectation value also cannot go out of the range of the distribution, as shown 
in Fig.~\ref{fig1} (left).

To enlarge the range of the distribution,
we combine the simulation data obtained at $\kappa=0.1400$, 0.1425 and 0.1440 
using the multipoint reweighting method explained in the previous section.
The red curve in the left panel of Fig.~\ref{fig1} is the result of the multipoint reweighting method.
We find that the red curve smoothly connects all the direct simulation results with small error bars. 
In the right panel of Fig.~\ref{fig1}, histograms from the multipoint reweighting method are plotted for $\kappa=0.1412$, 0.1424, 0.1436, and 0.144. 

The method is applicable to other observables.
The expectation values of 
$\partial S/ \partial \beta$ and 
\[[\partial S/ \partial \kappa]_{\rm SUB} \equiv \partial S/ \partial \kappa -(288 N_{\rm f} \kappa^4/c_0) (\partial S/ \partial \beta)\]
are needed in the calculation of the equation of state in the integral method.
In the right and left panels of Fig.~\ref{fig2}, we show the $\beta$- and $\kappa$-dependence of the 2-dimensional histogram of 
these quantities using the multipoint reweighting method.
The histogram moves as $\beta$ and $\kappa$ are varied.
We can compute the expectation values of $\partial S/ \partial \beta$ and $[\partial S/ \partial \kappa]_{\rm SUB}$ 
as continuous functions of 
$\beta$ and $\kappa$ without the overlap problem.
Once we obtain physical quantities as continuous functions of coupling parameters, 
we can calculate the lines of constant physics in the coupling parameter space
as well as the beta functions.

%%%%%%%%%%%%%%%%%%%%%%%%%%%%%%%%%%%%%%%%%%
\section{Lines of constant physics and beta functions}
\label{sec:beta-func}

\begin{figure}[tb]
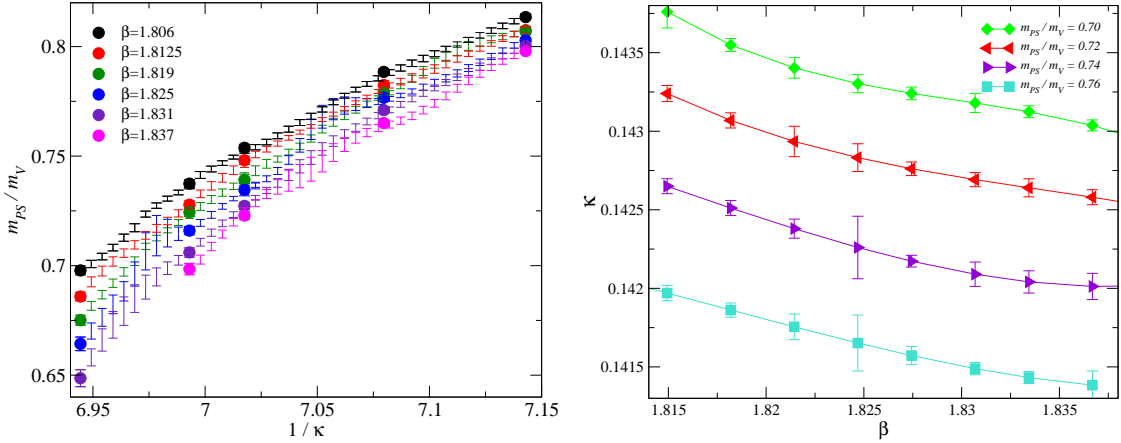

\begin{center}
%\vspace{-5mm}
\centerline{
\includegraphics[width=72mm]{pOVERrmesons_5K_II_PoS_2.eps}
\hspace{3mm}
\includegraphics[width=72mm]{POR_900x700_PoS_2.eps}
}
%\vspace{0mm}
\caption{Left: The pseudoscalar and vector meson mass ratio $m_{\rm PS}/m_{\rm V}$.
Right: The lines of constant physics for each $m_{\rm PS}/m_{\rm V}$
in the $(\beta, \kappa)$ plane.}
\label{fig3}
\end{center}
\vspace{-5mm}
\end{figure}

\begin{figure}[tb]
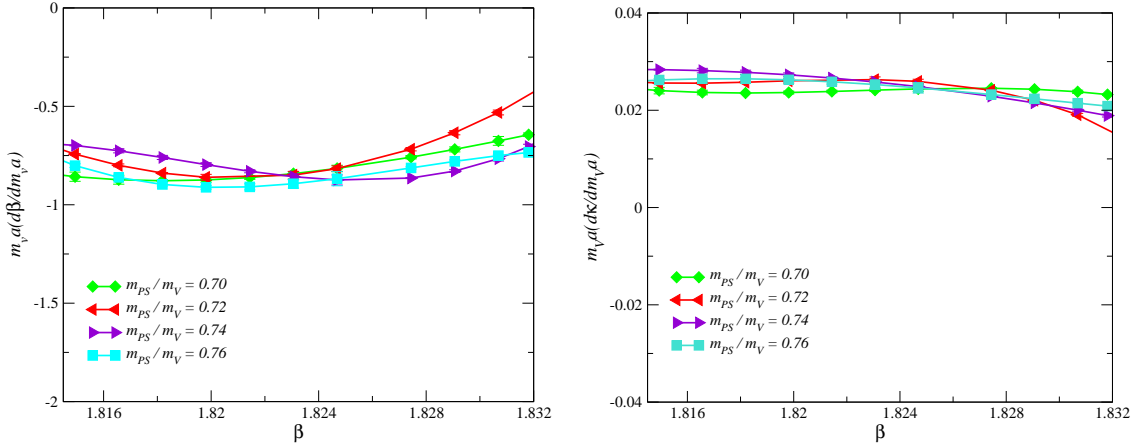

\begin{center}
%\vspace{-5mm}
\centerline{
\includegraphics[width=73mm]{derbeta_derRHO_PoS_2.eps}
\hspace{2mm}
\includegraphics[width=73mm]{derkappa_derRHO_PoS_2.eps}
}
%\vspace{0mm}
\caption{The beta functions: 
$(m_{\rm V} a) \partial \beta /\partial (m_{\rm V} a)$ (left) and 
$(m_{\rm V} a) \partial \kappa /\partial (m_{\rm V} a)$ (right).}
\label{fig4}
\end{center}
\vspace{-5mm}
\end{figure}

In this study, we define the lines of constant physics by fixing the dimension-less ratio of pseudoscalar and vector meson masses
$m_{\rm PS}/m_{\rm V}$ in the $(\beta,\kappa)$ space. 
On the lattice, we measure dimension-less observables $m_{\rm PS} a$ and $m_{\rm V} a$, where $a$ is the lattice spacing.
Because these lattice observables vary as we change $\beta$ or $\kappa$, 
$a$ is varied when we move along a line of constant physics.
The beta functions $\partial \beta /\partial a$ and $\partial \kappa /\partial a$ are defined 
through the variation $a$ along a line of constant physics.
The beta functions are needed in the calculation of the equation of state.

The multipoint reweighting method is useful for the calculation of the beta functions because we can calculate observables as continuous functions of $\beta$ and $\kappa$.
Combining the data at 30 simulation points (6 $\beta$'s $\times$ 5 $\kappa$'s) on the $16^4$ lattice
by the multipoint reweighting method, 
we obtain the mass ratio $m_{\rm PS}/m_{\rm V}$ plotted in the left panel of Fig.~\ref{fig3}.
From these data, determine the lines of constant physics for $m_{\rm PS}/m_{\rm V}= 0.70$, 0.72, 0.74 and 0.76,
as shown in Fig.~\ref{fig3} (right).
We then calculate the derivatives, $(m_{\rm V} a) \partial \beta /\partial (m_{\rm V} a)$ and 
$(m_{\rm V} a) \partial \kappa /\partial (m_{\rm V} a)$along each line of constant physics.
The results are shown in Fig.~\ref{fig4}.
Combining these beta functions with the measurement of $\langle \partial S/ \partial \kappa \rangle$ and 
$\langle \partial S/ \partial \beta \rangle$ on finite-temperature lattice, 
we can calculate the equation of state.

%%%%%%%%%%%%%%%%%%%%%%%%%%%%%%%%%%%%%%%%%%
\section{Conclusions and outlook}
\label{sec:conclusion}

We discussed the multipoint reweighting method in a multi-dimensional parameter space to avoid the overlap problem.
Using the method, we can reliably calculate histograms of physical quantities 
as well as the expectation values of the physical quantities, as continuous functions of coupling parameters.
These enable us to compute the lines of constant physics and the beta functions, 
which are needed in a calculation of the equation of state.

Our final objective is a study of finite density QCD. 
Using the multipoint reweighing method, we may absorb the main effect of the chemical potential by a change of $\beta$ and $\kappa$. 
If so, we may investigate finite density QCD avoiding the sign problem.

\paragraph{Acknowledgments}
This work is in part supported by Grants-in-Aid of the Japanese Ministry of Education, Culture, Sports, Science and Technology (Nos.\ 26400244, 26400251).


\begin{thebibliography}{99}
\bibitem{whot14} 
H.~Saito, S.~Ejiri, S.~Aoki, K.~Kanaya, Y.~Nakagawa, H.~Ohno, 
K.~Okuno, T.~Umeda (WHOT-QCD Collaboration),  
Phys.~Rev.~D {\bf 89}, 014508 (2014).
%arXiv:1309.2445.

\bibitem{FS89}
A.M.~Ferrenberg and R.H.~Swendsen,
Phys.~Rev.~Lett. {\bf 63}, 1195 (1989). 

\bibitem{CswBeta}
B.B.~Brandt, H.~Wittig, O.~Philipsen and L.~Zeidlewicz,
\pos{PoS(LATTICE 2010)172}.

\bibitem{whot10}
S.~Ejiri, Y.~Maezawa, N.~Ukita, S.~Aoki, T.~Hatsuda, N.~Ishii, K.~Kanaya, and T.~Umeda 
(WHOT-QCD Collaboration), 
Phys.~Rev.~D {\bf 82}, 014508 (2010). 
%[{\tt arXiv:0909.2121}].


\end{thebibliography}
\end{document}